\documentstyle[sprocl,epsfig]{article}

\def\Journal#1#2#3#4{{#1} {\bf #2}, #3 (#4)}

\def\NPB{{\em Nucl. Phys.} B}
\def\NPPS{{\em Nucl. Phys.} B {\em Proc. Suppl.}}
\def\PLB{{\em Phys. Lett.}  B}
\def\PRL{\em Phys. Rev. Lett.}
\def\PRD{{\em Phys. Rev.} D}
\def\PRe{\em Phys. Rep.}

\def\AnP{\em Ann. Phys. (NY)}
\def\JETP{\em JETP Lett.}

\def\beq{\begin{equation}}
\def\eeq{\end{equation}}

\def\fd{\rightarrow}

\def\ts{\left(}
\def\td{\right)}
\def\qs{\left[}
\def\qd{\right]}
\def\esp{\mbox{e}}

\def\bg{\beta}

\def\dg{\delta}

\def\lg{\lambda}

\def\sg{\sigma}

\def\Dg{\Delta}

\def\Og{\Omega}
\def\Phu{\Phi^{(1)}}
\def\Phd{\Phi^{(2)}}
\def\Phux{\Phu (x)}
\def\Phdx{\Phd (x)}
\def\Phuxp{\Phu(x)'}
\def\Phdxp{\Phd(x)'}
\def\Phdxs{\Phd(x)"}

\def\phu{\phi^{(1)}}
\def\phd{\phi^{(2)}}
\def\phux{\phu (x)}
\def\phdx{\phd (x)}
\def\phuxp{\phu(x)'}
\def\phdxp{\phd(x)'}
\def\phdxs{\phd(x)"}
\def\Tr{\mbox{Tr}}
\def\esp{\mbox{e}}

\def\be{\begin{equation}}
\def\ee{\end{equation}}
\def\bea{\begin{eqnarray}}
\def\eea{\end{eqnarray}}

\begin{document}

\title{A study of center vortices in $SU(2)$ and $SU(3)$
  gauge theories
\footnote{presented by M. Pepe.}}

\author{Michele Pepe}

\address{Inst. f\"ur Theoretische Physik, ETH H\"onggerberg, CH-8093
  Z\"urich, Switzerland\\
e--mail: pepe@itp.phys.ethz.ch}

\author{Philippe de Forcrand}

\address{Inst. f\"ur Theoretische Physik, ETH H\"onggerberg, CH-8093
  Z\"urich, Switzerland\\
and\\
CERN, Theory Division, CH-1211 Gen\`eve 23, Switzerland\\
e--mail: forcrand@itp.phys.ethz.ch} 


\maketitle\abstracts{
We show how center vortices and Abelian monopoles both appear as
local gauge ambiguities in the Laplacian Center gauge. Numerical
results, for $SU(2)$ and $SU(3)$, support the view that the string tension 
obtained in the center-projected theory matches the full string tension when
the continuum limit is taken.
}


\section{Introduction}
There are several approaches addressing the problem of understanding
the mechanism of color confinement in non-Abelian gauge theories. The
most popular share the idea that only a subset of the
degrees of freedom are relevant for
confinement. In the Abelian projection approach\cite{thooft1,mandel}, 
one takes into
account the maximal Abelian subgroup $U(1)^{N-1}$ of the gauge group
$SU(N)$ and monopoles are the effective degrees of
freedom under study. In the Center projection
approach\cite{thooft2,mack}, it is the center group
$Z_N$ which is considered and center vortices are the effective
degrees of freedom under study. 

These two schemes are commonly believed to give alternative
descriptions of confinement. Many analytical and numerical studies
have been and are being performed using these two approaches. They give clear
evidence that both Abelian and center degrees of freedom play a relevant
role. 

The reduction of the gauge symmetry and the selection of the effective 
degrees of freedom is usually carried out by a partial fixing of the
initial $SU(N)$ gauge freedom. We show that, in the Laplacian Center gauge,
monopoles and center vortices are closely related in a unified
description and that the latter are the effective degrees of freedom
relevant for confinement. Indeed center vortices are related to a more 
reduced gauge symmetry than monopoles. A fundamental issue of this
study is the use of the Laplacian Center gauge, which is free from the problem 
of the lattice Gribov copies affecting the widely used Maximal Center gauge.

\section{The center degrees of freedom}
Let us consider an $SU(2)$ gauge field defined in a plane and
expressed in terms of the radial and angular components 
$(A_r , A_\varphi )$. Suppose that the radial component is vanishing,
$A_r =0$, and the angular one is given by 
$A_\varphi =\frac{1}{2r} \sigma _3$. This gauge field configuration
describes a magnetic flux tube crossing the plane at the origin. 
If we consider a Wilson loop encircling the origin, we see that it
has a non--trivial value $\esp ^{i\pi\sigma_3} =-1$ with respect to the center
group $Z_2$ of $SU(2)$. Conversely a Wilson loop non encircling the
origin has a trivial value $+1$ with respect to $Z_2$. This is an
example of what a center vortex is. 

Consider now an $SU(2)$ gauge
theory on the lattice and decompose the link variable $U_{\mu} (x)$ as 
the product of two parts
\beq\label{decomp}
U_{\mu} (x) = Z_{\mu} (x) \, U_{\mu}' (x) 
\eeq
where $Z_{\mu} (x)$ lives in the center group $Z_2$ and $U_{\mu}' (x)$
is the coset part belonging to $SU(2)/Z_2$. For instance, this
splitting can be carried out defining $U_{\mu}' (x)$ as having
positive trace. If $W(C)$ is a Wilson loop
along the closed path $C$, making use of the decomposition 
(\ref{decomp}), we can write
\beq\label{WC}
W(C)=\sigma (C) \, W' (C) = 
\qs \prod_{p\in \Sigma}\, \sigma (p) \qd W' (C) 
\eeq
$W' (C)$ and $\sigma (C) \equiv \prod_{p\in \Sigma}\, \sigma (p)$ are
respectively the Wilson loops evaluated with the coset links 
$U_{\mu}' (x)$ and with the center links $Z_{\mu} (x)$. 
$\prod_{p\in \Sigma}$ is the product over all the plaquettes $p$
belonging to a surface $\Sigma$ supported by $C$; the value of 
$\sigma (C)$ does not depend on the choice of $\Sigma$ and, for
simplicity, we can choose it as the planar surface bounded by 
$C$. If we fix a gauge where 
$U_{\mu}'$ is smooth, then $W(C)$ has a non--trivial value with
respect to $Z_2$ if $\sigma (C)$ does. Moreover, considering 
(\ref{WC}) for a single plaquette, $\sigma (p) =-1$ is a ``signal'' for
a center vortex. In 4--dimensional space--time, center vortices form
closed surfaces in the dual lattice. 
\bigskip 

It is possible to give a
handwaving and qualitative argument showing how the described center
vortices can give rise to an area law behavior for the Wilson
loops. According to the previous argument, we are interested in the
plaquettes $p$ having $\sigma (p) =-1$. Suppose that $t$ is the
probability for a plaquette to have $\sigma (p) =-1$, consequently
$(1-t)$ is the probability to have $\sigma (p) =+1$. Consider a
time--slice of a lattice $SU(2)$ gauge configuration, then center
vortices form 1--dimensional closed strings in the dual lattice. The
following figure displays an example of such a time--slice

\begin{figure}[h]
\begin{center}
\psfig{figure=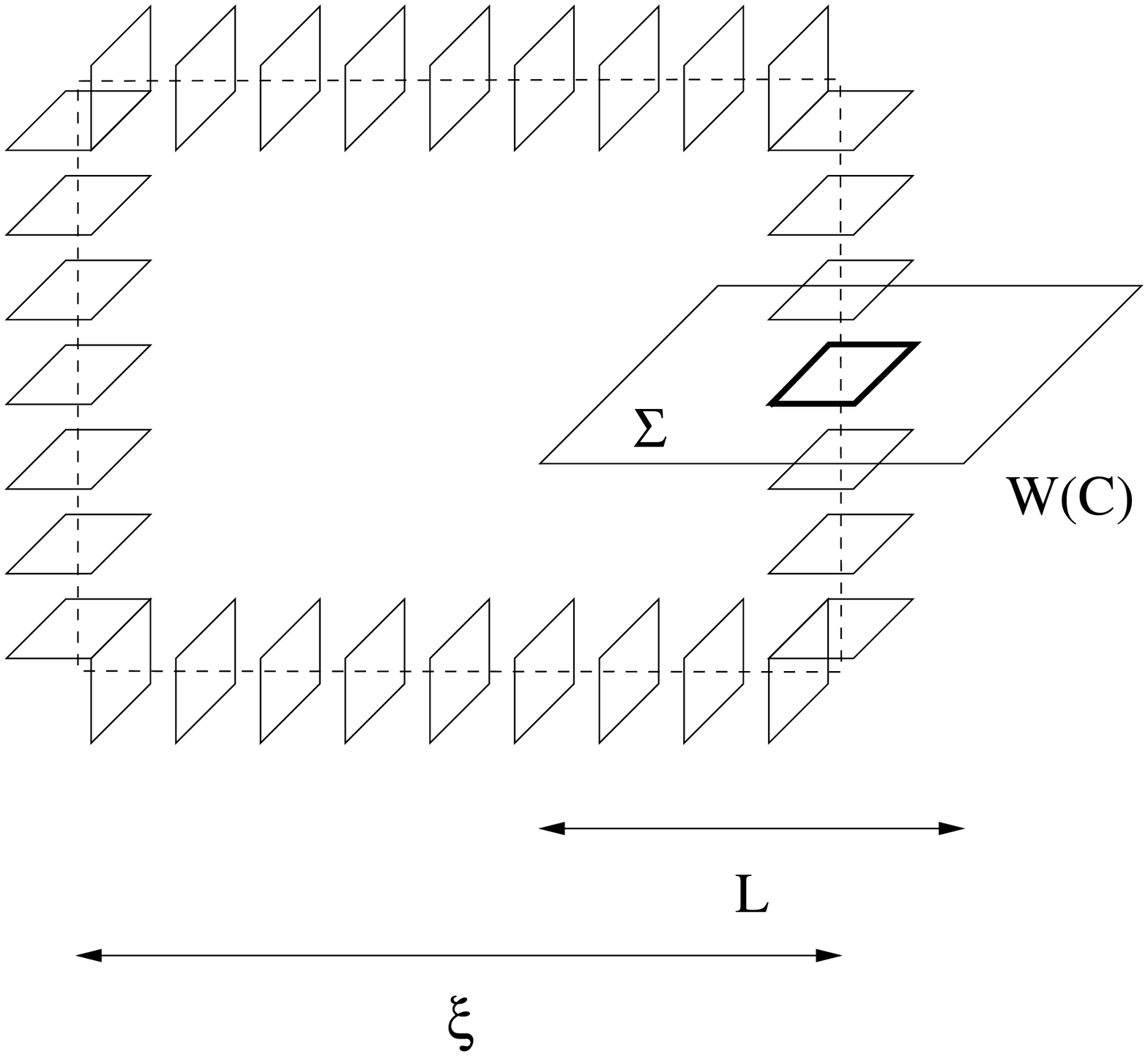,height=1.5in}
\end{center}
\end{figure}

The drawn plaquettes have $\sigma (p) =-1$ and the broken line is the
closed string in the dual lattice representing the center
vortex. Consider a Wilson loop $W(C)$ which, in the
figure, is represented by the continuous line; the plaquette in bold line
belongs to the planar surface $\Sigma$ spanned by $C$. Let $L$
be the linear extension of $C$ and $\xi$ the average linear size of
the center vortex strings. If $\xi \gg L$, the value $\sigma (p)$ 
of each plaquette $p$ in $\Sigma$ is independent of the others; then,
with respect to the center degrees of freedom, one can write
\beq
< W(C) > \sim <\sigma (p) > ^A = \esp^{A\log (1-2t)}
\eeq
where $A$ is the area of $\Sigma$. With this qualitative argument we
have obtained on one hand that the center degrees of freedom can give
rise to an area law behavior for the Wilson loop and, on the other
hand, that the string tension is about twice the probability $t$ for a
plaquette to have $\sigma (p) =-1$. We have assumed that $\xi \gg L$: 
in order for this inequality to be satisfied for arbitrarily large Wilson loops,
$\xi$ must be divergent, that is center vortex strings must percolate through the
lattice. Conversely, suppose that the center vortices do not percolate,
so that $\xi$ has a finite value, then for $L$ sufficiently
large, one must have $\xi \ll L$. The next figure displays such a
case

\begin{figure}[h]
\begin{center}
\psfig{figure=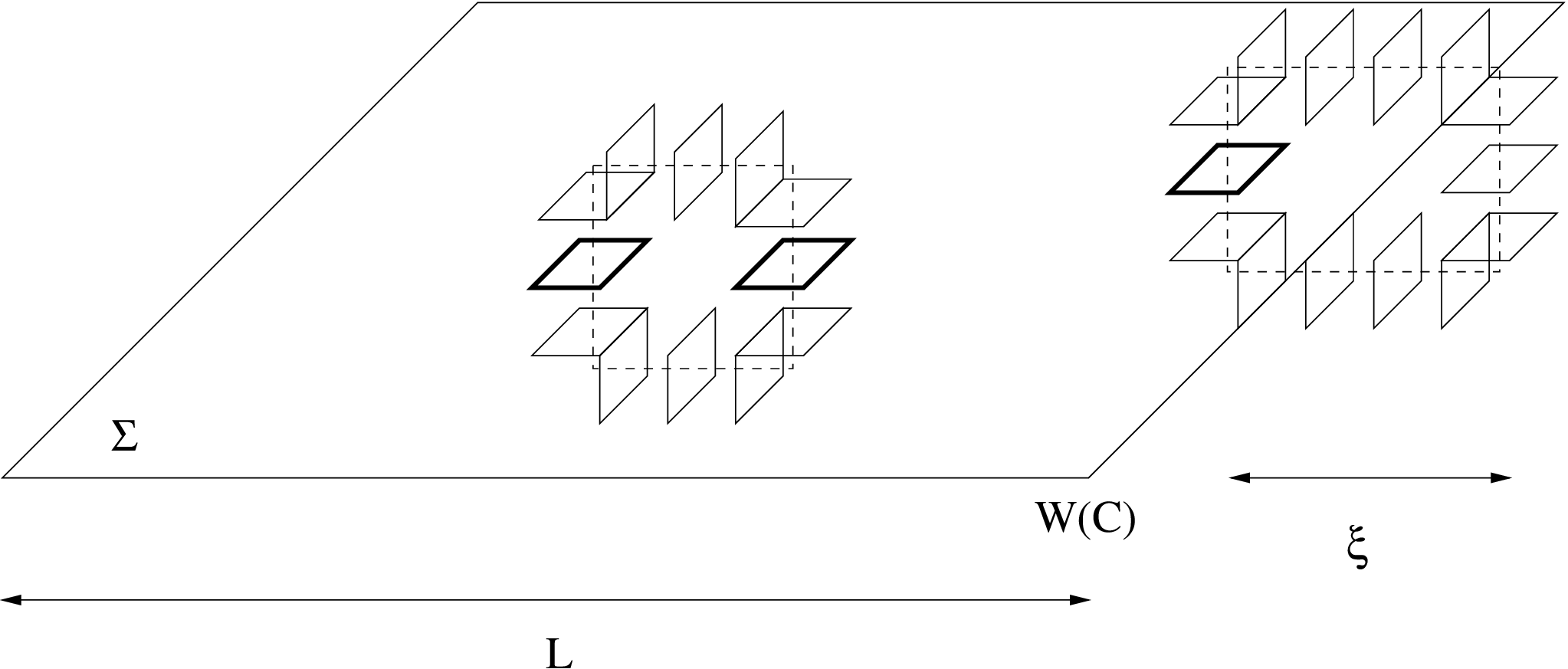,height=1.5in}
\end{center}
\end{figure}

The values $\sigma (p)$ of the plaquettes in $\Sigma$
well inside the boundary $C$ are strongly correlated by pairs; moreover
these pairs do not 
give a net contribution to $\sigma (C)$. So only the plaquettes
belonging to a ring of thickness $\xi$ around the boundary of
$\Sigma$ can give a non--trivial contribution. According to this
reasoning and, with respect to the center degrees of freedom, we can
write
\beq
< W(C) > \sim <\sigma (p) > ^{\xi P} = \esp^{\xi P\log (1-2t) }
\eeq
where $P$ is the perimeter of $\Sigma$. Thus, we have obtained a perimeter
law behavior for the Wilson loop. The conclusion from this handwaving 
argument is that the center degrees of freedom can give rise to an
area law for the Wilson loop and that the area/perimeter behavior can 
be recast in terms of percolation/non-percolation of center
vortices.

\section{Laplacian gauge fixing}
Many lattice simulations -- starting with the initial results by
Greensite and collaborators\cite{green1} -- have been performed in
order to investigate
the role of the center degrees of freedom in the non perturbative
features of the non-Abelian gauge theories. In these studies the
selection of the center degrees of freedom is carried out by the numerical 
partial gauge fixing of an ensemble of configurations.
The most widely used gauge is the Maximal
Center gauge. In this gauge, it has been shown\cite{phmax1,phmax2}
that the removal of the
center vortices leads to the loss of the area law for the
Wilson loop, to the restoration of chiral symmetry and to the disappearance
of non--trivial topological features. 
The numerical implementation
of the Maximal Center gauge fixing is performed by a local iterative
procedure and, as 
there are many local extrema, lattice Gribov copies are present.
This is a serious problem\cite{K&T,borny} that 
can lead to a complete loss of meaningful information in the
$Z_2$ projected model. Thus, it is important to study the role of the
center degrees of freedom in the confinement mechanism 
considering a smooth gauge not affected by
this problem of the lattice Gribov copies. In the Laplacian gauge the
reduction of the gauge degrees of freedom is carried out in an
unambiguous way. This gauge was proposed by Vink and Wiese\cite{vink}: they
suggested  to use the eigenvectors of the
covariant Laplacian operator to fix the gauge.
Since we are interested in reducing unambiguously the symmetry of the gauge
group from $SU(N)$ to its center $Z_N$, it is
useful to consider the Laplacian operator in the adjoint
representation. In fact, as the adjoint representation is invariant
under gauge transformations in $Z_N$, the adjoint Laplacian
procedure fixes unambiguously the gauge up to the center symmetry.

Consider the 4--dimensional lattice $SU(N)$ gauge
theory. The adjoint covariant Laplacian operator $\Dg^{ab}_{yx} (\dot{U})$ 
is given by
\beq\label{adjlap}
-\Dg^{ab}_{yx} (\dot{U}) = \sum_{\mu} 
\ts 2\,\dg_{y,x}\,\dg^{ab} - \dot{U}_{\mu}^{ab} (x-\hat{\mu}) \,\dg_{y,x-\hat{\mu}}
- \dot{U}_{\mu}^{ba} (x) \,\dg_{y,x+\hat{\mu}}\td
\eeq
where $a$,$b = 1,\ldots,(N^2-1)$ are color indices and $x$,$y$ are
space--time lattice coordinates. The dotted $\dot{U}_{\mu} (x)$
are the link variables in the adjoint representation
and are related to the links $U_{\mu} (x)$ in the fundamental by
\beq\label{adjlink}
\dot{U}_{\mu}^{ab} (x) = \frac{1}{2}\Tr 
\ts \lg _a U_{\mu} (x) \lg _b U_{\mu}^{\dagger} (x) \td
\eeq
$\lg_i$, $i=1,\ldots ,(N^2-1)$ being the generators of $SU(N)$  with the
normalization $\Tr (\lg _a \lg_b)=2\dg_{ab}$. 
If $V$ is the
volume of the lattice, $\Dg (\dot{U})$ is a $[(N^2-1)V]\times [(N^2-1)V]$ real
symmetric matrix which depends on the gauge field. 
The eigenvalues $\mu_j$ of $\Dg$ are real and the eigenvector equation is
\beq\label{eigeq}
\Dg^{ab}_{yx} (\dot{U}) \phi^{(j)} _b (x) = \mu_j \,\phi^{(j)} _a (y) 
\eeq
where $\phi^{(j)}$, $j=1,\ldots ,[(N^2-1)V]$ are the
(real) eigenvectors. So we can associate $(N^2-1)$--dimensional real vectors
$\phi^{(j)} (x)$ to every lattice site $x$. 
Consider now a gauge transformation on the links in the
fundamental representation 
$U_{\mu}' (x) = \Og (x) U_{\mu} (x) \Og^{\dagger} (x+\hat{\mu})$;
the eigenvector
equation (\ref{eigeq}) becomes
\beq
\dot{\Og}^{\dagger}\,^{ai} (y) \Dg^{ik}_{yx} (\dot{U}')  \dot{\Og}^{kb} (x)
\phi^{(j)} _b (x) =  \mu_j \,\phi^{(j)} _a (y)
\eeq
where $\dot{\Og}$ is the gauge transformation in the adjoint
representation. 
This relation shows that the eigenvalues are gauge invariant and the
eigenvectors transform according to 
$\dot{\Og}^{ab} (x) \phi^{(j)} _b (x) = \phi^{(j)}_a(x)' $. This transformation 
law can be rewritten as follows
\beq\label{trlaw}
\Og (x) \Phi^{(j)} (x) \Og^{\dagger} (x) = \Phi^{(j)}(x)'
\eeq
where we have defined the $su(N)$ matrices -- i.e. in the $SU(N)$
algebra --
$\Phi^{(j)} (x)=\sum_{a=1}^{N^2-1} \phi^{(j)}_a (x) \lg_a$ and
$\Phi^{(j)}(x)'=\sum_{a=1}^{N^2-1} \phi^{(j)}_a(x)' \lg_a$.
Gauge transformations rotate the vectors $\phi^{(j)} (x)$ in color space and 
we can fix the gauge by requiring a conventional arbitrary orientation 
for the $\phi^{(j)} (x)$. 
To perform the reduction of the gauge symmetry from $SU(N)$ to $Z_N$ we only need 
to fix the orientation of two eigenvectors of the Laplacian operator.
As we are interested in
the non--perturbative features and in fixing a smooth gauge, we
consider the two lowest-lying eigenmodes $\phi^{(1)}$ and $\phi^{(2)}$. 

The gauge fixing procedure can be split into two steps. In the first, 
one rotates $\Phux$ at every $x$ so that $\Phuxp$ is diagonal.
This leaves a residual symmetry corresponding to gauge transformations
in the Cartan subgroup $U(1)^{N-1}$.  
To further reduce the gauge freedom we have to consider a 
second step where the second eigenvector $\phi^{(2)}$ is taken into
account. The gauge transformation that has rotated $\Phux$ to the Cartan 
subalgebra, maps $\Phdx$ to $\Phdxp$. 
The remaining $U(1)^{N-1}$ symmetry can be fixed to $Z_N$ by requiring that some
conventionally chosen color components of the twice rotated matrix
$\Phdxs$ vanish.
Now we describe explicitly how to perform the
presented two--step program for the $SU(2)$ gauge theory\cite{phmax2}, where, at
each $x$, we consider the 3--dimensional real vectors $\phux$ and
$\phdx$.

{\bf{Step 1}}. We consider equation (\ref{trlaw}) for the first eigenvector and we
conventionally define $\Og (x)$ to be the gauge transformation that
rotates $\phux$ along the direction 3 in color space:
$\dot{\Og} (x) \phux = \phuxp \propto (0,0,1)$.
$\Og (x)$ is unambiguously defined up to gauge transformations $V(x)$ 
in $U(1)$. So we have reduced the gauge symmetry to the Cartan
subgroup of $SU(2)$. This is the Laplacian Abelian gauge\cite{vds1,vds2}.

{\bf{Step 2}}. We apply the gauge transformation $\Og (x)$ found in {\bf{Step 1}} to
$\phdx$: $\phdxp = \dot{\Og} (x) \phdx$. $\phdxp$ is not invariant
under the rotations $V(x) \in U(1)$ that leave  $\phuxp$ unchanged; then
we can fix this symmetry by requiring, for example, that 
$\phdxs = \dot{V} (x) \phdxp$ lie in the 1--3 color plane in the
positive direction.

We have completely fixed the gauge symmetry in the adjoint
representation and, as it is center--blind, we are left with the
center symmetry $Z_2$. The described two--step procedure for $SU(2)$
can be extended\cite{noi} to $SU(N)$ with the same pattern of gauge symmetry
fixing: $SU(N) \fd U(1)^{N-1}$ in the first step and 
$U(1)^{N-1} \fd Z_N$ in the second one.

Two last, important remarks concern the accidental degeneracy of
$\mu_1$ and $\mu_2$ in (\ref{eigeq}) and the arbitrariness in the
eigenvectors $\phi^{(j)}$. 
If it happens that $\mu_1$ or $\mu_2$ is degenerate, 
the gauge fixing can not be carried out unambiguously and one has 
a (global) Gribov ambiguity. This case is really exceptional and never
occurs in the numerical simulations. 
The second point is about the scale and sign arbitrariness of the
eigenvectors $\phi^{(j)}$. Rescaling can not give rise to any
ambiguity in the procedure while the freedom in the choice  
of the sign does. This global freedom can be eliminated with a conventional
choice on $\phi^{(j)}$.

\section{Local gauge ambiguities}
The adjoint Laplacian gauge fixing procedure has local 
defects. Now we discuss how these defects can show up and how they
can be identified with monopoles and center vortices in $SU(2)$. This
discussion can be generalized to $SU(N)$\cite{noi}.

{\bf{Step 2 ill--defined}}: the second step of the Laplacian gauge
fixing is not defined at the points $x$ where $\phux \, // \, \phdx$. 
In such a case also $\phdxp$ is invariant under rotations 
$V(x) \in U(1)$. The condition $\phux \, // \, \phdx$ sets two
constraints  and so these
points $x$ --  where the gauge symmetry is promoted from $Z_2$ to
$U(1)$ -- form 2--dimensional surfaces in 4--dimensional space--time.

{\bf{Step 1 ill--defined}}: the gauge fixing procedure can not be even 
started at the points $x$ where $\phux =(0,0,0)$. These defects
constitute 1--dimensional strings in the 4--dimensional space--time since
3 constraints must be satisfied. At these points the symmetry is not
fixed and the gauge freedom is $SU(2)$.

The 2--dimensional surfaces of {\bf{Step 2 ill--defined}} can be
identified with center vortices. Suppose that at a point $x_0$ 
it happens that $\phu (x_0) \, // \, \phd (x_0)$, then moving along a small
loop around the singularity point $x_0$, $\phd$ describes a $2\pi$
rotation in color space. As a $2\pi$ phase in the adjoint
representation of $SU(2)$ corresponds to a $\pi$ phase in the
fundamental one, it follows that a Wilson loop encircling $x_0$ has a
non trivial value with respect to the center $Z_2$. 

The 1--dimensional strings of {\bf{Step 1 ill--defined}} can be
identified with monopole world--lines. In fact, in analogy with the
Georgi--Glashow model, the points $x$ where the Higgs field vanishes
and the gauge symmetry can not be reduced from $SU(2)$ to $U(1)$,
correspond to the monopole world--lines. 

In the adjoint Laplacian
gauge, monopoles and center vortices turn out to be closely related
in a unified description. Consider the 2--dimensional
surface ${\cal{S}}$ of the center vortices. At every 
$x\in {\cal{S}}$, $\phux$ and $\phdx$ are parallel or anti--parallel. So
${\cal{S}}$ is divided in parts where $\phux$ is parallel to $\phdx$ and
parts where it is anti--parallel. By continuity, 
these patches must be separated by 1--dimensional strings where 
$\phux =0$ or $\phdx =0$. Moreover, in the neighbourhood of a
monopole, $\phux$ has a hedgehog--like shape and  there will be
a direction where $\phu \, // \, \phd$. Thus monopole world--lines are
embedded within the 2--dimensional surfaces of center vortices. 

\section{Numerical results and their interpretation}
We have performed numerical simulations to investigate the role of the 
center degrees of freedom in the $SU(2)$ and $SU(3)$ lattice gauge
theories. For $SU(2)$ we have collected 1000 configurations at 
$\beta=2.3$, $2.4$, $2.5$ on a $16^4$ lattice; for $SU(3)$ we have
generated 500 configurations on a $16^4$ lattice at $\beta=6.0$. The
following figure shows the measurement of the Creutz ratios 
$\chi(R)=-\ln (\langle W(R,R)\rangle \langle W(R-1,R-1)\rangle 
/\langle W(R,R-1)\rangle^2)$ for $SU(2)$ at $\beta=2.4$ ($W(R,T)$ is
the $R\times T$ Wilson loop). 
\begin{figure}[h]
\begin{center}
\psfig{figure=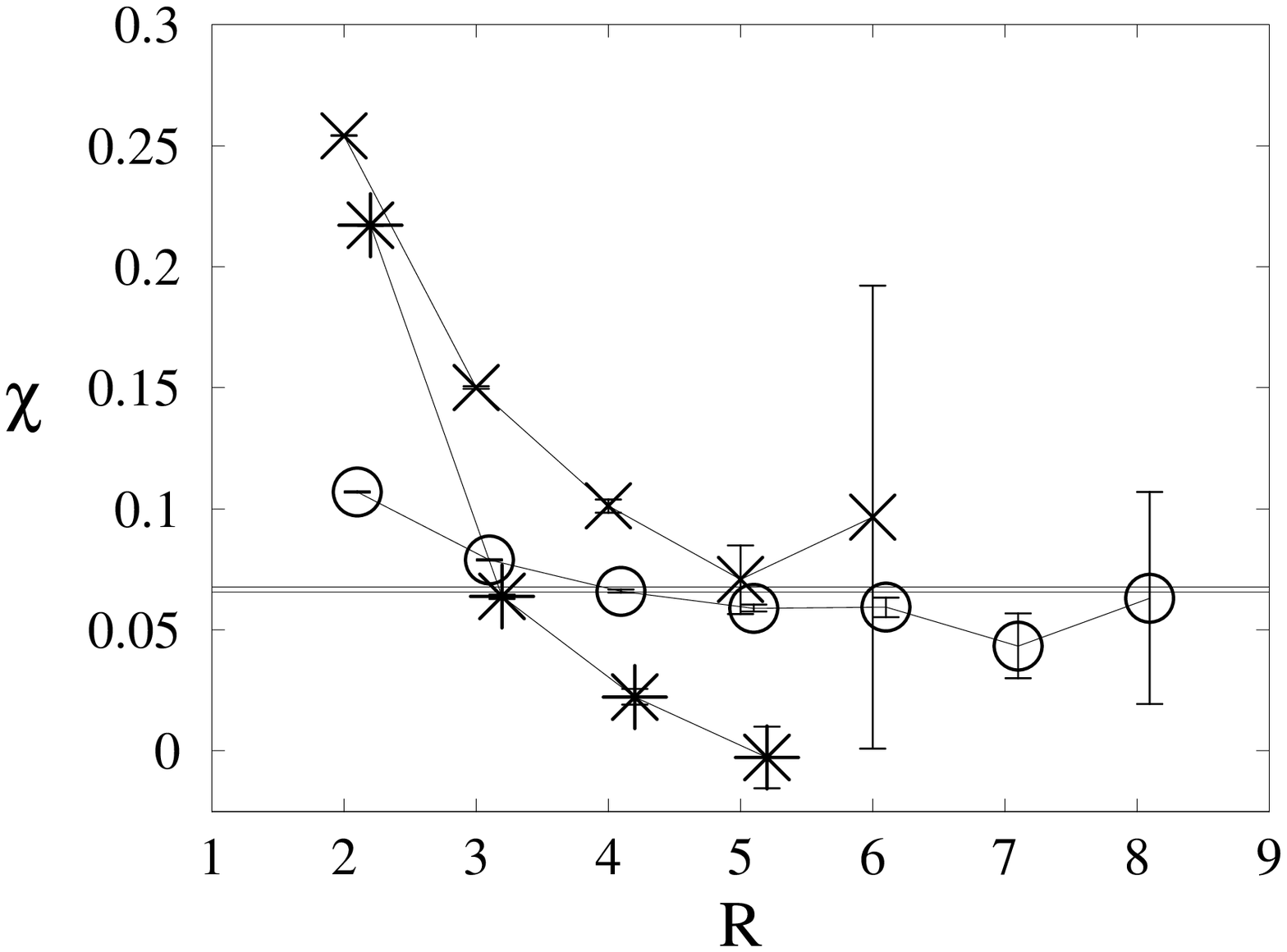,height=1.5in}
\end{center}
\end{figure}
Crosses refer to $SU(2)$, circles to center projection after Laplacian
gauge fixing and stars to the coset part. The continuous band is the
value in the literature\cite{teper,bali1} for 
the $SU(2)$ string tension at the considered set of
parameters. The results show, on one hand, the flattening of 
the Creutz ratios in the $Z_2$ sector and, on the other hand, the
vanishing of the Creutz ratios computed with the coset links. We have
obtained a similar behaviour for $\beta =2.3$ 
and $2.5$. In the case of $SU(3)$ also, the following figure shows the  
Creutz ratios in the $Z_3$ sector after Laplacian gauge fixing.
\begin{figure}[h]
\begin{center}
\psfig{figure=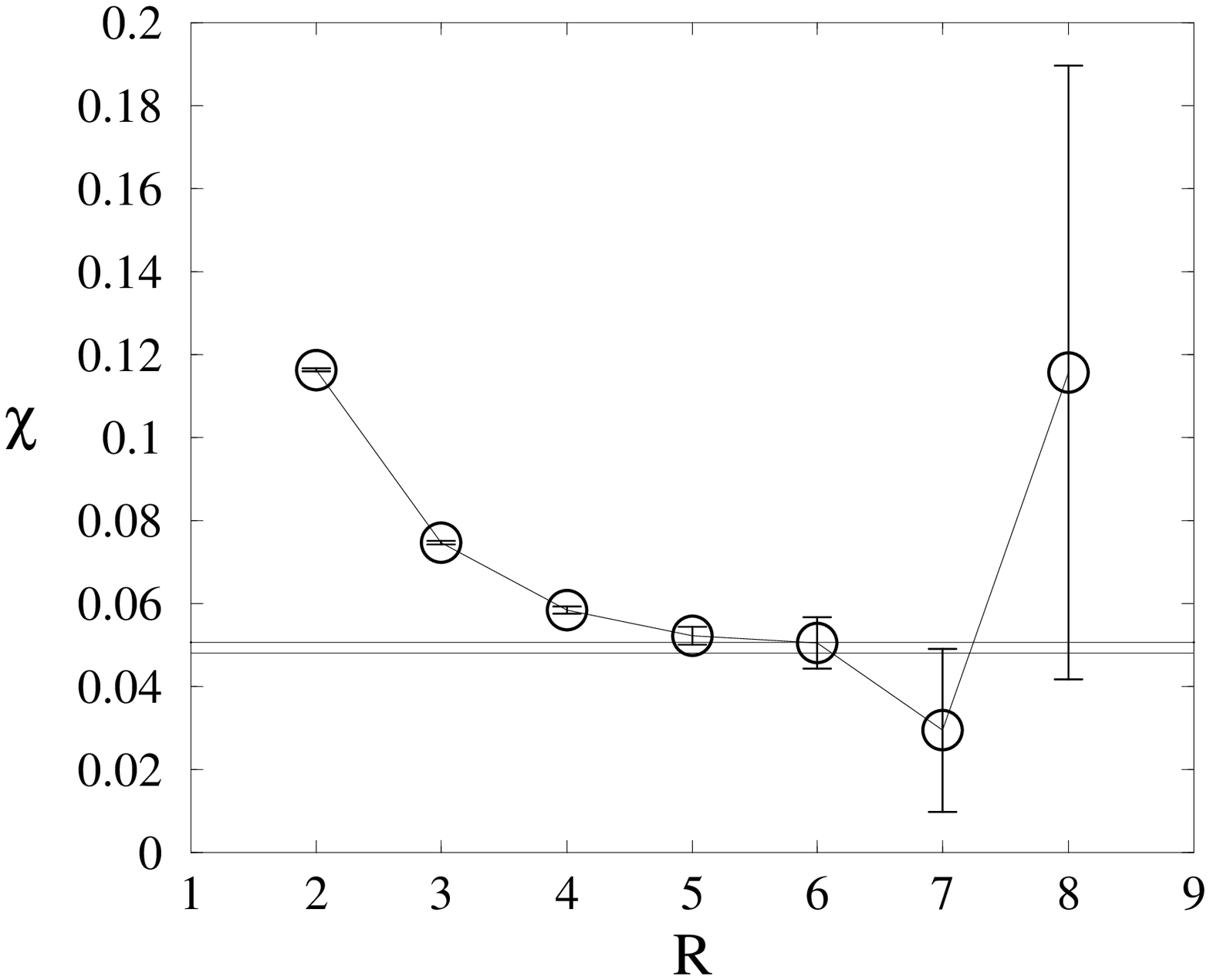,height=1.5in}
\end{center}
\end{figure}
The continuous band is the value in literature\cite{bali2} for the 
$SU(3)$ string tension at the chosen set of parameters. Also in
this case, one can clearly see flattening to a non vanishing
value for the Creutz ratios evaluated with center projected~links. 

The good agreement with the values in the literature for
the string tension in $SU(2)$ and $SU(3)$ should not be
over--estimated. Numerical simulations are performed at finite
lattice spacing and lattice artifacts can give
non-negligible effects in the center projected theory. Our conjecture is
that even if, at finite lattice spacing, the flattening value of the
Creutz ratios in the center sector changes with the particular lattice 
Laplacian used to fix the gauge, this dependence 
vanishes in the continuum limit. The observation of such a behaviour
would be a robust confirmation of the relevance of the center degrees
of freedom in the confinement mechanism. To investigate this
issue, we have considered three different lattice Laplacians differing 
by irrelevant operators: they have been built using smeared links in
(\ref{adjlink}).
Every set of 1000 configurations at $\beta=2.3$,
$2.4$, $2.5$, has been fixed in each one of the three gauges and
the Creutz ratios have been measured after center projection. The
table summarizes our results:
\begin{table}[h]
\begin{center}
\footnotesize
\begin{tabular}{||c||c|c|c||}
\hline
{} & $\bg =2.3$ & $\bg =2.4$ & $\bg =2.5$ \\
\hline
$R_0$ & $0.813(23)$ & $0.860(20)$ & $0.978(18)$\\
$R_1$ & $0.592(12)$ & $0.720(11)$ & $0.804(12)$\\
$R_2$ & $0.547(8) $ & $0.653(7) $ & $0.739(11)$\\
\hline
\end{tabular}
\end{center}
\end{table}
$R_i =\sqrt{\sg_i/\sg_{SU(2)}}$ where $\sg_i$ is the string tension measured
in the $Z_2$ sector;
$i=0,1,2$ is an index for the three lattice
Laplacians and $\sg_{SU(2)}$ is the value in literature for the 
$SU(2)$ string tension at the three values of $\bg$. Thus,
it is in the continuum limit ($\bg\fd \infty$) that the
string tension measured after center projection correctly reproduces 
the value of the full gauge theory.

\section*{Acknowledgments}
We thank  C. Alexandrou, M. D'Elia, S. D\"urr and J. Fr\"ohlich for
useful discussions and G. Bali for providing us with a code for data analysis.

\end{document}